\documentclass[12pt]{article}
\usepackage{amstex,epsfig}
\begin{document}

\begin{center}
 
{\bf Nonleptonic Hyperon Decays with QCD Sum Rules}
\end{center}
\vspace{.5in}
 
\noindent E.M. Henley \\
\noindent 
{\em Department of Physics and Institute for Nuclear Theory, \\
Box 351560, University of Washington, Seattle, Washington 98195}\\

\vspace{2mm}
\noindent W-Y.P. Hwang \\

\noindent {\em Department of Physics, National Taiwan University, \\
Taipei 106, Taiwan } \\
 
\vspace{2mm}
\noindent L.S. Kisslinger \\

\noindent {\em Department of Physics, Carnegie-Mellon University, \\
Pittsburgh, Pennsylvania 15213}
 
\vspace{.5in}
{\bf Abstract}
 
Despite measurements which date more than 20 years ago, no straightforward
solution of the ratio of the  parity-conserving (P-wave) to parity-
violating  (S-wave) decays of the hyperons has been obtained. Here we use
two 2-point methods in QCD sum rules to examine the problem. We find that 
resonance contributions are needed to fit the data, similar to a
chiral perturbation theory treatment.
\vspace{.5in}
 
\noindent
PACS Indices: 12.15.Ji, 13.30.Eg, 12.38.Lg, 11.50.Li
 
\newpage
 
\section{Introduction}

The nonleptonic decays of the hyperons occur with pion emission, e.g.,
$\Lambda^0 \rightarrow p + \pi^-$. Measurements of the decay rates and the S/P
(pv/pc) ratios of the emitted pions were carried out over 20 years ago 
\cite {DGH}. They remain of interest today because no one has been able 
to provide a relatively simple explanation of the S/P ratios.  

To-date a variety of approaches have been used. Some of the early work used a
soft pion approach  \cite {DGH}. In this limit the $\Sigma^+ \rightarrow n
\pi^+$ decay with an S-wave pion vanishes. The $\Sigma^- \rightarrow n \pi^-$
decay amplitude can be obtained approximately by an adjustment of the SU(3) F/D
ratio. The soft pion approach in the S-wave and poles in the P-waves approach 
(see Fig.1) was used by Donoghue et al.\cite {DGH}, who argue that there
could also be a direct coupling, as shown in Fig. 1d, but they too have
difficulty in fitting the S/P ratios. Other work is that of ref. \cite{Z}.
Most recently, Barasoy and Holstein
\cite {BH} have used chiral perturbation theory, but have had to include
$(70, 1^-)\frac{1}{2}$ resonances and parameters to obtain a
reasonable fit to the data. 

Of the seven decays, $\Sigma^+  \rightarrow p \pi^0,~ \Sigma^+ \rightarrow n
\pi^+,~ \Sigma^- \rightarrow n \pi^-,~
\Lambda^0 \rightarrow n \pi^0,~ \Lambda^0 \rightarrow p \pi^-, ~\Xi^-
\rightarrow \Lambda^0 \pi^-,~ \Xi^0 \rightarrow \Lambda^0 \pi^0$, there
are only
four independent ones if isospin symmetry holds. Experimentally, the SU(3) 
27-plet is smaller than the octet by a factor of approximately 20. Like those
before us, we choose the 4 independent decays as those with a charged pion,
namely\\

\begin{center}
\begin{eqnarray}
\Sigma^+_+ &:& \ \ \Sigma ^+ \rightarrow n \pi^+ , \nonumber \\
\Sigma^-_- &:& \ \  \Sigma^-\rightarrow n \pi^-, \nonumber \\
\Lambda^0_- &:& \ \ \Lambda^0 \rightarrow p \pi^-, \\
\Xi^-_- &:& \ \ \Xi^- \rightarrow \Lambda^0 \pi^-.\nonumber
\end{eqnarray}
\end{center}

%\vspace {2 in}

   In the present work we use the method of QCD sum rules with two 2-point
formulations for the three-point correlators needed to obtain coupling
constants, which we discuss in the next section.   We find that in order 
to find stable solutions for the sum rules we must explicitly introduce
single-pole resonance contributons, analogous to the addition of resonance
contributions in Ref. \cite{BH}. Since this introduces new constants which
can, however, be used to fit the data, 
we did not proceed to investigate the last two decays
$(\Lambda^0_- \,$ and $ \Xi^-_-)$.
\begin{figure}
\begin{center}
\epsfig{file=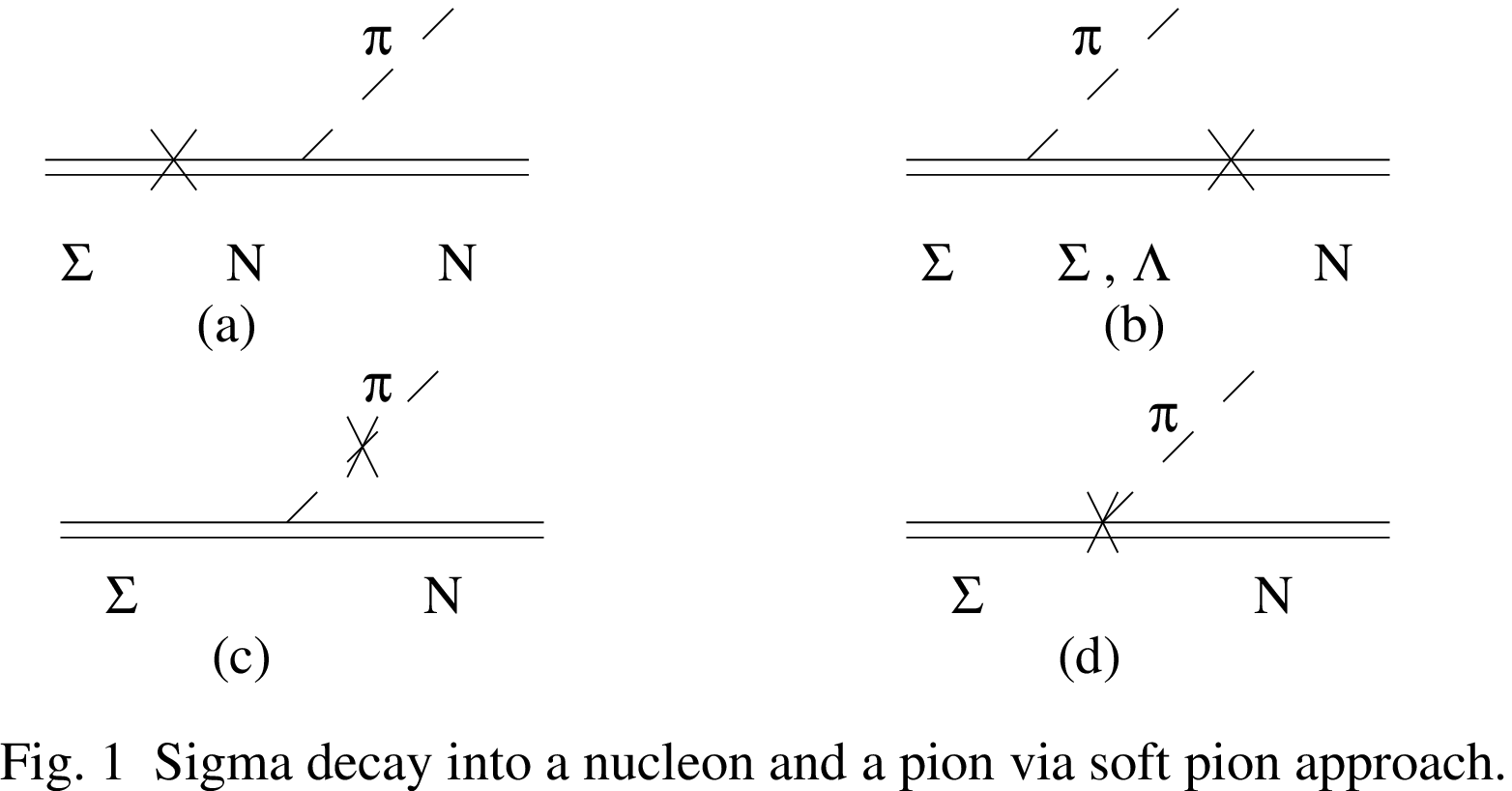,width=10cm}
%caption{}
{\label{Fig.1}}
\end{center}
\end{figure}

\section {Methodology}

\subsection {QCD Sum Rules}

QCD sum rules were introduced by Shifman, Vainshtein, and Zhakarov
\cite{SVZ}.  It is a useful method to obtain properties of hadrons and it
uses QCD explicitly.
The short range perturbative QCD is extended by an operator product
expansion (OPE)
of the correlator, giving a series in 
inverse powers of the squared momentum with Wilson coefficients.  
The  convergence at low momentum is improved by using a type of Laplace
transform, called the Borel transform.  The coefficients 
involve universal quark and gluon condensates.  This quark--based calculation 
of a given correlator is equated to the same correlator obtained via a 
dispersion relation, giving sum rules from which a property
can be estimated.   
The method can be extended for quantities in an external field, 
such as the magnetic coupling to a nucleon in an electromagnetic field
\cite{ISBK}.

%\vspace {2 in}
%Fig. 1

The method begins with a correlator

\begin{equation}
\Pi (p) = i \int d^4 x \; e^{iq \cdot x} < 0 \mid T [\eta (x) \bar{\eta} 
(0)] \mid 0 > \; ,
\end{equation}

\noindent where $\eta$ has the quantum numbers of the hadron being studied.  
For a proton, we may take

\begin{eqnarray}
\label{eta}
\eta (x) &=& \epsilon^{abc} [u^{aT} (x) C \gamma_\mu u^b (x)] \gamma^5 
\gamma^\mu d^c (x)\; , \\
\bar{\eta} &=& \epsilon^{abc} [\bar{u}^b \gamma_\nu C \bar{u}^{aT}] \bar{d}^c 
\gamma^\nu \gamma^5 \; ,\nonumber
\end{eqnarray}

\noindent where $a,b,c$ are color indices
and the notation of Bjorken and Drell is 
used.  The quark field
operators $d,u,s$ destroy these quarks, $C$ stands for charge 
conjugation and $T$ for transpose.

The ``currents" $\eta$ are not unique\cite{L}, but the form given in
Eq. (\ref{eta})
has been used by many authors.
The correlator can be written as an operator product expansion

\begin{equation}
\Pi = C_s I  + \sum_n C_n (p^2) O_n
\end{equation}

\noindent where the operators $O_n$ can be ordered by dimension and the 
corresponding Wilson coefficients decrease by increasing powers of $p^2$.

The correlators $\Pi$ have structure functions $\Pi^j$, each of which 
satisfies a dispersion relation $(P^2 = -p^2)$

\begin{equation}
\Pi^j (P^2) = \frac{1}{\pi} \int_0^\infty \frac{Im \Pi^j (s) ds}{s+P^2}\;
\end{equation}

Subtraction terms in Eq.(4) are eliminated by means of a Borel transform, 
which guarantees convergence,

\begin{eqnarray}
B [F (p^2) ] &=& 
\lim_{\stackrel{n \rightarrow\infty}
{-p^2 \rightarrow\infty \: , \: -p^2/n \rightarrow M_B^2}}
(-p^2)^{(n+1)} ( \frac{d}{d(-p^2)} )^n F(p^2) \\
&=& \frac{1}{\pi} \int_0^\infty ds \; Im F(s) e^{-s/ M_B^2}\; . \nonumber
\end{eqnarray}

There are a number of ways to use these sum rules:

(i) a two-point method with or without an external field;

(ii) a three-point method with couplings and momentum transfers considered 
explicitly.  This method has fewer susceptibilities but it is more 
complicated; it may require non--local condensates.

In this article we will use only two-point methods.  We will compare the
two point method in an external field with the two point method with a pion
creation matrix element. 

\subsection{$\Sigma^-_- $ in an External Pion Field}

The calculation of the non-leptonic decays of hyperons is similar to that of 
the weak pion-nucleon coupling constant. If we neglect the mass difference 
in the baryon octet, then $\Sigma \rightarrow N \pi$ is quite akin to 
$N \stackrel{weak}{\rightarrow} N \pi$.
The primary difference is that the latter is due to weak neutral currents 
and the former due to charged currents.

We use the operators

\begin{eqnarray}
\eta_{\Sigma^-}  &=& \epsilon^{abc} [d^{aT} C \gamma_\mu d^b] \gamma^5 
\gamma^\mu s^c \; ,\\
\eta_n &=& \epsilon^{abc} [\bar{d}^b \gamma_\nu C \bar{d}^{aT}] u^c 
\gamma^\nu \gamma^5 \nonumber
\end{eqnarray}

\noindent for the $\Sigma^-$ and for the $n$. In addition, we need 
the weak interaction, for which we use the local one,

\begin {eqnarray}
H_W &=&\frac{G_F}{\sqrt{2}}J^\mu J_\mu^\dagger\; , \nonumber\\
J^\mu &=& \bar{u} \gamma^\mu (1-\gamma_5)s \;sin\theta_C
+ \bar{u}\gamma^\mu(1-\gamma_5)d \;cos\theta_C \; ,
\end{eqnarray}
where $G_F$ is the Fermi coupling constant and $\theta_C$ is the Cabibbo
angle. The correlator is 
\begin{eqnarray}
\Pi = i \int d^4x \; e^{ip \cdot x} <0 \mid T[\eta_\Sigma(x)\, H_W \, \bar
{\eta}_n(x)] \mid 0>_\pi,
\end{eqnarray}
\noindent where the quarks propagate in an external field. Since the
point at which the
external pion field is at zero momentum transfer\cite{ISBK},
the three-point function for the vertex is reduced to a two-point function.
To the order considered here, the QCD diagrams which contribute 
to the correlator are shown in Fig.2.
%\vspace {2 in}
%Fig. 2
\begin{figure}
\begin{center}
\epsfig{file=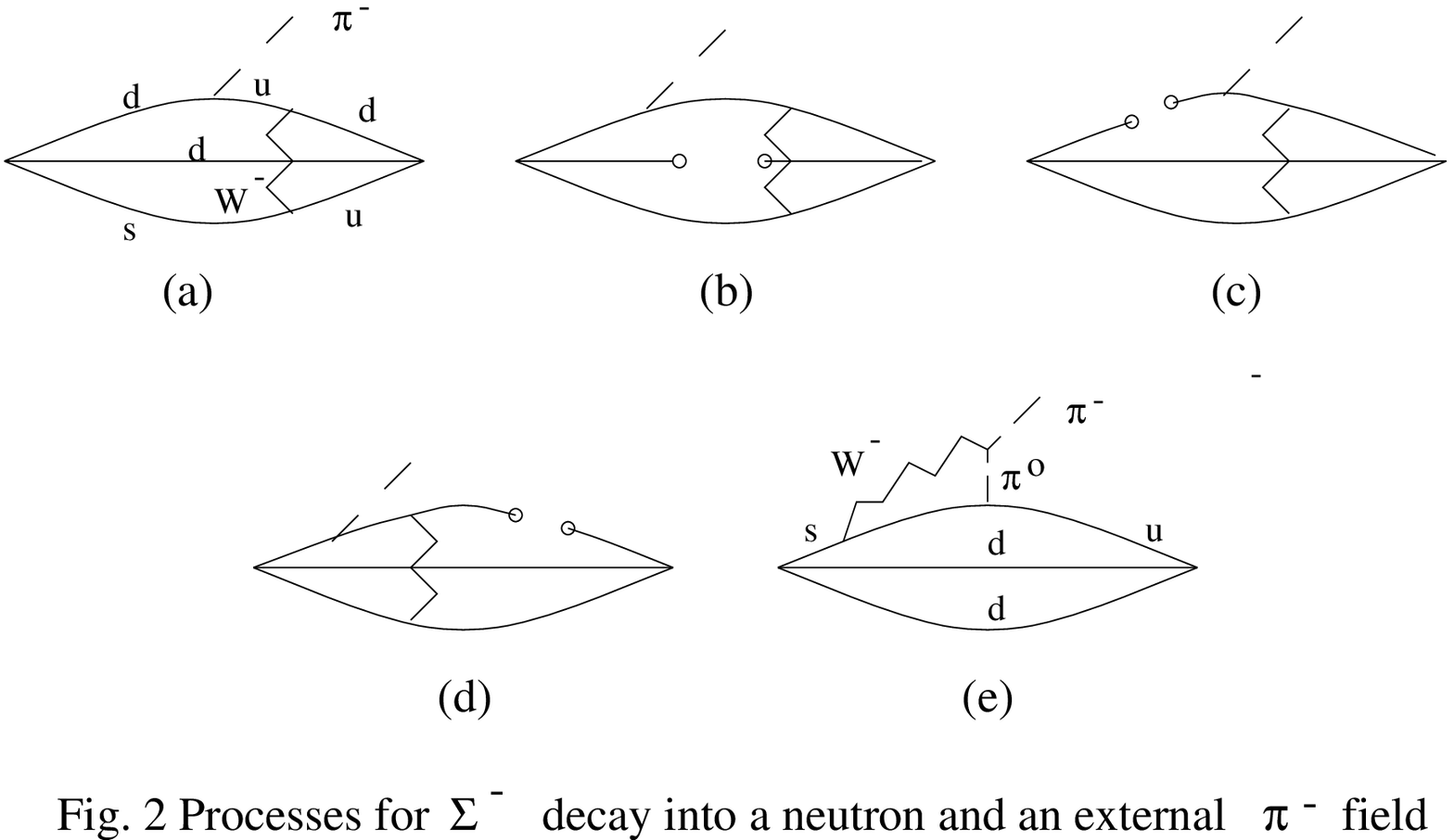,width=10cm}
%caption{}
{\label{Fig.2}}
\end{center}
\end{figure}
In the diagrams, 
the wavy line represents a $W^\pm$ boson, the dashed line represents a 
pion. The diagrams are evaluated in momentum space.  
Fig. (2a) gives no contribution.  Diagram (2e) is quite 
different from the others.  It involves the weak matrix element 
\begin{equation}
<\pi^- \mid J_\alpha \mid \pi^0 >= \sqrt{2} F_\pi q_\alpha \; ,
\end{equation}
where $F_\pi$ is the 
weak pion form factor and $q$ is the momentum of the $\pi^0$.  
The contribution of this diagram cannot be neglected.  There are additional 
higher order terms, e.g., gluonic corrections, which we omit here. Reasonjs
will become apparent further on. 

We obtain the following results, using dimensional regularization in $4-
\epsilon $ dimensions
\begin {eqnarray}
Fig. 2b&:&  -\frac{16A <\bar{q} q> p^2 \ln{ p^2}}{(4 \pi)^4 \epsilon} [\not{p}
(1 -
\gamma \epsilon + \frac{43}{24}\epsilon -\frac{\epsilon}{2} \ln{p^2}) + 4m
(1 - \gamma \epsilon + \frac{15}{8}\epsilon - \frac{\epsilon}{2} \ln{P^2})]
\nonumber\\
(1 - \gamma_5)\nonumber \; ,\\
Fig. 2c&:& \frac{4A <\bar{q} q> p^2 \ln{ p^2}}{(4 \pi)^4 \epsilon}[\not{p}(1 -
\gamma \epsilon + \frac{8}{3}\epsilon -\frac{\epsilon}{2} \ln{p^2}) + 4m
(1 - \gamma \epsilon + \frac{11}{4}\epsilon - \frac{\epsilon}{2} \ln{p^2})]
\nonumber\\
(1 - \gamma_5) \nonumber \;\\
Fig. 2d&:&  \frac{4A <\bar{q} q> p^2 \ln{ p^2}}{(4 \pi)^4}[ \frac{5}{9}\not{p}
 - \frac{13}{4 \epsilon}m
(1 - \gamma \epsilon + \frac{355}{156}\epsilon - \frac{13\epsilon}{8} 
\ln{p^2})](1 - \gamma_5) \; ,\\
Fig. 2e&:& \frac{\sqrt{2} F_\pi p^4A \ln{ p^2}}{(4 \pi)^6 6}[p^2 +
\frac{6 m \not{p}}{\epsilon}(1 +\frac{7}{2}
\gamma \epsilon + \frac{\epsilon}{6} -\frac {9}{8}\epsilon \ln{p^2})]
(1 + \gamma_5) \; , \nonumber
\end{eqnarray}
\noindent
where $\gamma$ is the Euler constant, $m$ the strange quark mass,  and 
\begin {equation}
A= \sqrt{2} G_F \sin{\theta_C} \cos{\theta_C} \; .
\end{equation}
\noindent
From these equations, it is clear that, as in determining the weak pion
nucleon 
coupling constant \cite {HHK}, we need to include  vertex renormalizations. 
There are several of these, shown in Fig. 3.
%\vspace {2 in}
%Fig. 3
\begin{figure}
\begin{center}
\epsfig{file=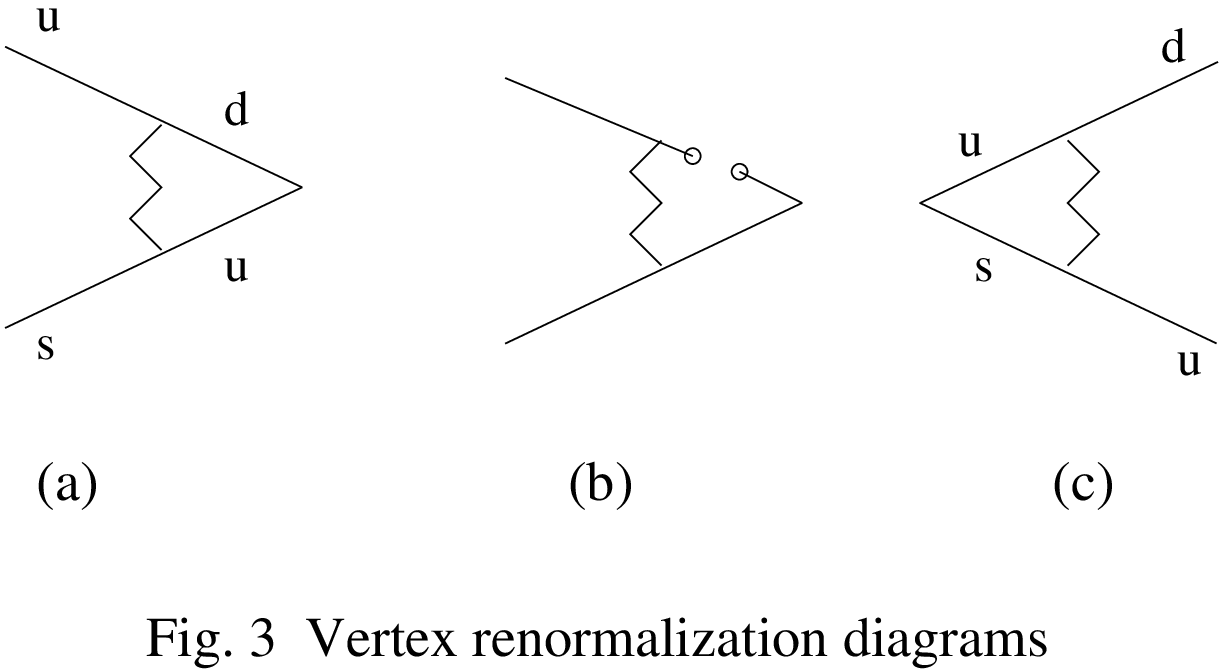,width=10cm}
%caption{}
{\label{Fig.3}}
\end{center}
\end{figure}
For Fig. 3a we obtain
\begin {eqnarray}
\bar {\eta}&=& \bar {u} \Gamma_\nu C \bar{s} \gamma^\nu \gamma_5 \;,\\
\Gamma_\nu(3a) &=& \frac{A}{(4 \pi)^2 \epsilon}(1 - \frac{\gamma
\epsilon}{2} + 
\frac{\epsilon}{2})\gamma_\nu (k^2)^{1- \epsilon /2} (1 - \gamma_5) \; .
\nonumber
\end{eqnarray}
For Fig 3b we get
\begin{eqnarray}
\Gamma_\nu(3b) = -\frac{A}{6} (1- \frac{\epsilon}{2})<\bar{q}q> \frac{\not{k}}
{k^2} \gamma _\nu (1 - \gamma_5) \; .
\end {eqnarray}
For Fig. 3c we obtain
\begin {eqnarray}
\eta &=& d^T C \gamma^\nu \Gamma_\nu u \;,\\
\Gamma_\nu(c) &=& - \frac{8Am}{(4 \pi)^2 \epsilon} \gamma_\nu (1 - \frac{\gamma
\epsilon}{2} + \frac{\epsilon}{2}) (k^2)^{- \epsilon/2} (1 - \gamma_5) \; .
\nonumber
\end {eqnarray}
Employing these 3 vertex renormalizations and calculating the 
diagrams of Fig.2 with them, we obtain 
\begin {eqnarray}
\Pi= -\frac{4A p^2 \ln{p^2}}{(4 \pi)^4} \{[\frac{35}{18}\not{p}
+\frac{19}{6}m]<\bar{q} q>(1 - \gamma_5)
-[ \sqrt{2}\frac{ F_\pi p^4}{24(4 \pi)^2}  + \ldots] (1 + \gamma_5)\}\; .
\end {eqnarray}
We have not carried out the renormalization for Fig. 2c because we shall see 
that it is not needed. 

For the phenomenological (or so-called right-hand) side we have
\begin{eqnarray}
\label{rhs}
\Pi = - \lambda_N \lambda_\Sigma \frac{1}{\not{p} - M_N}(A_S +A_P \gamma_5) 
\frac{1}{\not{p}-M_\Sigma}\nonumber\\
+ \frac{1}{\not{p} - M_N} (\bar{c}_1 + \bar{c}_2 \gamma_5) \frac{1}
 {\not{p} - M^*} \; ,
\end{eqnarray}
plus the continuum. The inclusion of resonances is indicated by the 
$(\bar{c}_1 + \bar{c}_2 \gamma_5)$ term; the mass $M^*$
represents the resonance mass. The resonance terms can be
separated into two parts, each of which is a single pole term; we use
$\frac{\bar{c}_a^- + \bar{c}_b^- \gamma_5}{M(\not{p} - \bar{M})}$ 
for these single pole terms. The choice of $\bar{M}$ is arbitrary; 
any other choice will simply lead to different values for 
$ \bar{c}_a^-$ and $\bar{c}_b^-$. The single-pole quantities 
$ \bar{c}_a^-,\bar{c}_b^-$ are functions of momentum, or after the Borel 
transform they are called $c_a^-,c_b^-$ and are functions of the Borel 
mass, $M_B$.
We define $\bar{M} \equiv \frac{1}{2}(M_N + M_\Sigma)$ and $\bigtriangleup M
\equiv M_\Sigma - M_N$. To first order in $\bigtriangleup M$ we find
\begin{eqnarray}
\Pi = - \frac{\lambda_N \lambda_\Sigma}{(p^2 - \bar{M}^2)^2}[A_S(p^2 +
\bar{M}^2) - A_P \gamma_5 (p^2 - \bar{M}^2) + A_S 2 \bar{M}\not{p}
+A_P \bigtriangleup M \not{p} \gamma_5] \nonumber\\ 
+ \lambda_N \lambda_\Sigma 
\frac{(\not{p} + \bar{M})(c_a^- + c_b^- \gamma_5)}{M(p^2 - \bar{M}^2)} \; .
\end {eqnarray}
We abandon the $ \not{p}$ and $ \not{p} \gamma_5$ sum rules because 
$A_P \bigtriangleup M$ would vanish in the analogous pion-nucleon vertex sum 
rule; this term is too sensitive to $\bigtriangleup M$. The single pole 
term  within the square bracket is similar to that which occurs
in determining the pion-nucleon coupling constant. Fig. 2e
requires no renormalization for the $p^2$ term; we have not carried out the
renormalization for the $m
\not{p}$ and $m \not{p} \gamma_5$ terms. After inclusion of the continuum
and carrying out a Borel transform we obtain
\begin{eqnarray}
\label{sum1}
\Pi &=& - \frac{A}{4^5 \pi^6 L^{4/9}}[\frac{76}{6}M_B^4 a m E_2 (1-\gamma_5) +
\sqrt{2}\frac {M_B^8 E_3}{4} F_\pi (1 + \gamma_5)] \nonumber \\
 &=&-\lambda_N \lambda_\Sigma[ A_S(\frac{2 \bar{M}^2}{M_B^2} -1) + A_P 
\gamma_5 + (c_a^- + c_b^- \gamma_5)] e^{- \frac{\bar{M}^2}{M_B^2}} \; ,
\end{eqnarray}
where $E_n$ represents the continuum contribution,
\begin{equation}
E_n = 1 - (1 + x + \frac{x^2}{2} +\ldots + \frac{x^n}{n!}) e^{-x}\; ,
\end{equation}
with $x = s/M_B^2$, where s is the continuum threshold. For the nucleon it 
was found that $ s \approx 2.3 GeV^2$ and for the $\Sigma' \, s \approx 3.2
GeV^2$ 
\cite {CPW};  
here we usually take an intermediate value of $s = 2.8 GeV^2$ for 
the transition, but explore other values for stability. $\lambda_N$
 and 
$\lambda_\Sigma$ are known from previous studies \cite {CPW}:
$\tilde{\lambda}_N
\tilde{\lambda}_\Sigma = (2 \pi)^4 \lambda_n \lambda_\Sigma =0.303 GeV^6 $\\

A crucial point in the analysis of the sum rules to to recognize that there
is some uncertainty in the p-dependence of the single-pole terms. The
assumption used in Eq.(\ref{sum1}) is the simplest possible. There almost
certainly is other $M_B$ dependence of the pole term, which makes no
difference at the point $M_B = \bar{M}$, but which can give stable
solutions. We find the reasonable but not obvious behavior that in every
case the single-pole constants are linear functons of $M_B$.

As for the pion-nucleon weak coupling constant, we gain some stability by 
multiplying both sides of the QCD sum rules by $M_B^2$ and carrying out
$(1- M_B^2 \partial / \partial M_B^2)M_B^2 ~\Pi$. This removes the great 
sensitivity
in $A_S$ to $M_B$ from the double pole
on the right-hand side, although some remains. We obtain
\begin{eqnarray}
A_S &=&- \frac{\frac{A}{4^3 \pi^2 L^{4/9}}(\frac{76}{3}M_B^6 a m E_2 +\sqrt{2}
E_4 M_B^{10} F_\pi) - \tilde{\lambda}_N \tilde{\lambda}_\Sigma c_a^- \bar{M}^2
e^{-M^2/M_B^2}
}{\tilde{\lambda}_N \tilde{\lambda}_\Sigma [2 (1-\frac{\bar{M}^2}
{M_B^2})+1]\bar{M}^2 e^{-\bar{M}^2/M_B^2}}\; ,\\ 
A_P &=-&~\frac{\frac{A}{4^3 \pi^2 L^{4/9}}(\frac{76}{3}M_B^6 a m E_2
-\sqrt{2} E_4 M_B^{10} F_\pi) +\tilde{\lambda}_N \tilde{\lambda}_\Sigma
c_b^- \bar{M}^2 e^{-M^2/M_B^2}
}{\tilde{\lambda}_N \tilde{\lambda}_\Sigma
\bar{M}^2 e^{-\bar{M}^2/ M_B^2}}.
\end{eqnarray}

\begin{figure}
\begin{center}
\epsfig{file=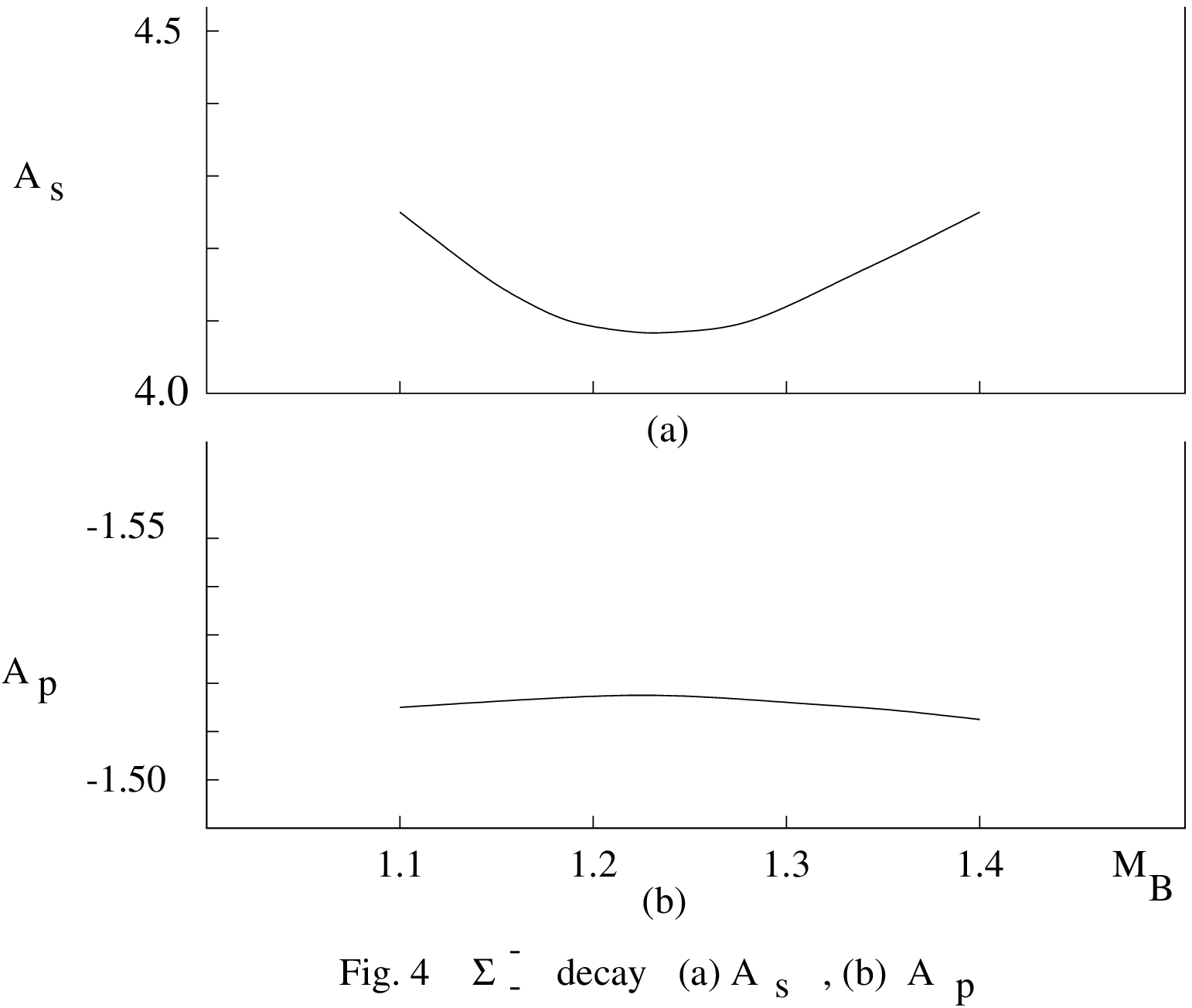,width=10cm}
%caption{}
{\label{Fig.4}}
\end{center}
\end{figure}

The parameters that used in this work are A = 3.57 $10^{-6}$ GeV$^{-2}$,
m=.15 GeV, a=.55 GeV$^3$, and L = 0.621 ln(10 M$_B$).
The experimental values are $A_S= (4.27 \pm 0.02)\times 10^{-7},
A_P= (-1.52 \pm 0.16)\times 10^{-7}, A_S/A_P \approx -2.8$. 
The single-pole parameters which give stable solutions for $A_s$ are \\ 
$c_a^-$ = (-6.37+10.7 $M_B$)$\times 10^{-7}$, and for $A_p$ are
 $c_b^-$ = (1.37 - .25 $M_B$)$\times 10^{-7}$. The
solutions are shown in Fig. 4.

\subsection {$\Sigma^+_+$ Decay in an External Field}

For $\Sigma^+$ we take
\begin{equation}
\eta_\Sigma^+  = \epsilon^{abc}[u^{aT} C \gamma_\mu u^b] \gamma^5 \gamma^\mu
s^c \; .
\end{equation}
The relevant diagrams are shown In Fig. 5
%\vspace {2 in}
%Fig. 5
\begin{eqnarray}
Fig. 5a&:&  0 \nonumber\\
Fig. 5b&:& -\frac{2A<\bar{q} q> p^2 \ln{p^2}}{(4 \pi)^4} (\frac {4}{3} \not {p}
+\frac{25}{6} m) (1 -\gamma_5) \nonumber\\
Fig. 5c&:& \frac{2A <\bar{q} q> p^2 \ln{p^2}}{(4 \pi)^4}(\frac {4}{3} \not{p} +
\frac {3}{2} m) (1 - \gamma_5)\nonumber\\
Fig. 5d&:&  0\nonumber\\
Fig. 5e&:&  0   \\
Fig. 5f&:& \frac{8 A p^4 \ln{p^2}}{(4 \pi)^6 \epsilon}[p^2(1 - \frac{3}{2} 
\gamma 
\epsilon + \frac{25}{6} \epsilon - \frac{9}{8} \epsilon \ln {p^2})
- 4 m \not{p} (1  - \frac{3}{2} \gamma \epsilon + \frac{73}{16} \epsilon
- \frac{9}{8} \epsilon \ln{p^2})] ( 1 + \gamma_5) \nonumber\\
Fig. 5g&:& -\frac{8 A <\bar{q} q> p^2 \ln{p^2}}{(4 \pi)^4 \epsilon}[2 \not{p}
(1 -\gamma \epsilon + \frac{3}{2} \epsilon- \frac{\epsilon}{2} \ln{p^2})
 + m (1 - \gamma \epsilon + \frac{\epsilon}{2} -\frac{\epsilon}{2} \ln{p^2})]
 (1 - \gamma_5) \nonumber\\
Fig. 5h&:& - \frac{4 A <\bar{q} q> p^2 \ln{p^2}}{(4 \pi)^4 \epsilon}[\not{p}
(1 -\gamma \epsilon + \frac{5}{4} \epsilon- \frac{\epsilon}{2} \ln{p^2})
+ 2 m (1 - \gamma \epsilon + \frac{3}{2}\epsilon -\frac{\epsilon}{2} \ln{p^2})]
 (1 - \gamma_5) \nonumber\\
Fig. 5i&:& 0 \nonumber \\
Fig. 5j&:&  \frac{4 A <\bar{s} s> p^2 \ln{p^2}}{(4 \pi)^4 \epsilon}[\not{p}
(1 -\gamma \epsilon + \frac{9}{4} \epsilon- \frac{\epsilon}{2} \ln{p^2})]
 (1 + \gamma_5) \nonumber
\end{eqnarray}
\begin{figure}
\begin{center}
\epsfig{file=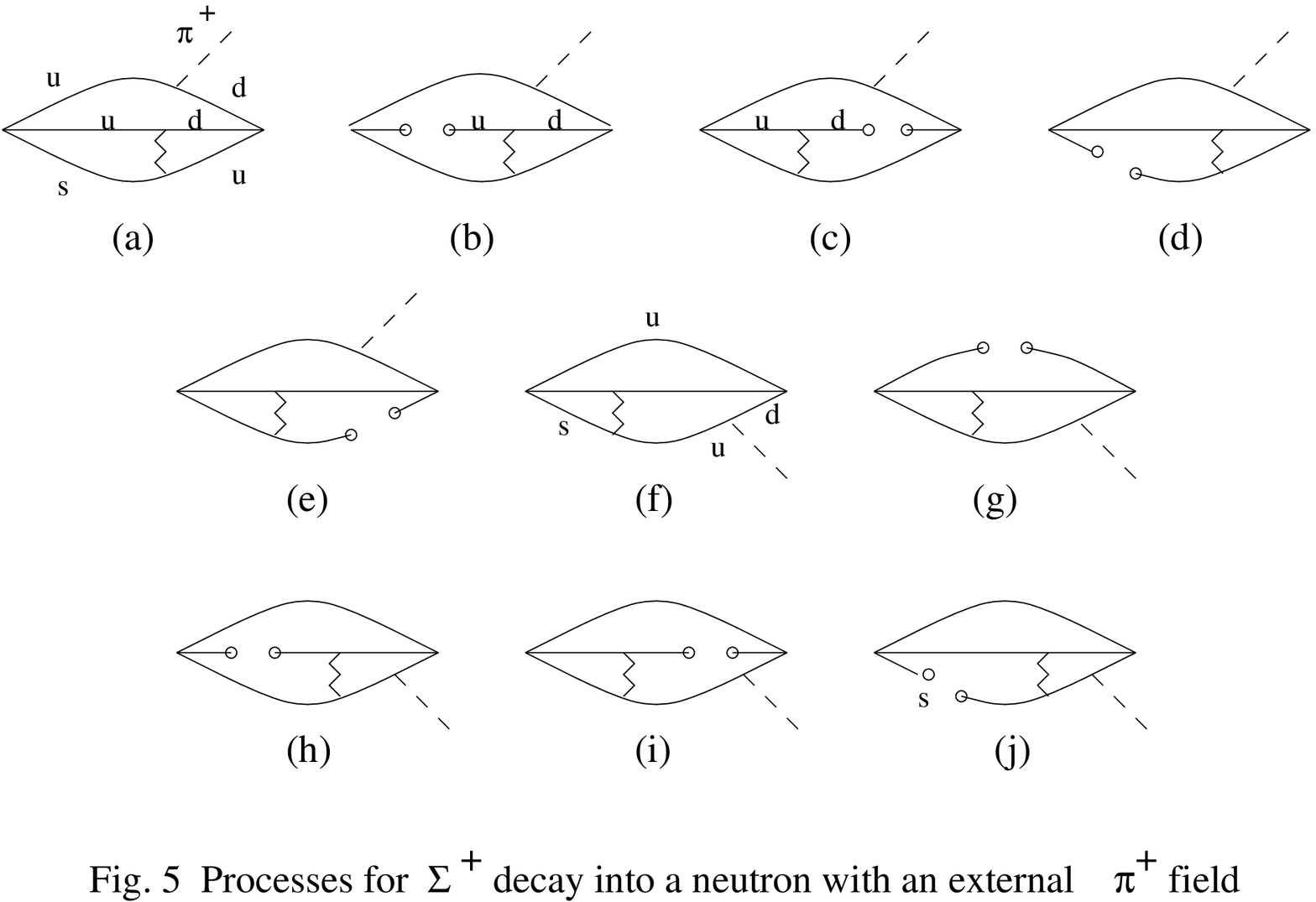,width=10cm}
%caption{}
{\label{Fig.5}}
\end{center}
\end{figure}
Once again we need to carry out vertex renormalizations for Figs. 5f, g, h and 
j. We omit the details and simply show the results. As for the $\Sigma^-$ 
decays we omit the $\not{p}$ and $\not{p} \gamma_5$ terms. We find

\begin{eqnarray}
\Pi = - \frac{10}{3} \frac{A m <\bar{q} q>p^2\ln{p^2}}{(4 \pi)^4}(1 - \gamma_5)
    +  8 (\frac{8}{3} - \frac{\gamma}{2})\frac {A p^6 \ln{p^2}}{(4 \pi)^6}
(1 + \gamma_5) \; .
\end {eqnarray}

After taking a Borel transform,  adding anomalous dimensions and 
continuum contributions, this becomes

\begin{eqnarray}
\Pi = \frac{10}{3} \frac{A m <\bar{q} q>}{(4 \pi)^4}(1 - \gamma_5)M_B^4 E_2
L^{-4/9}
- 48 (\frac{8}{3} - \frac{\gamma}{2})\frac {A }{(4 \pi)^6}
(1 + \gamma_5) M_B^8 E_4 L^{-4/9}\; .
\end {eqnarray}

\begin{eqnarray}
A_S &=& -\frac{\frac{A}{4^3 \pi^2 L^{4/9}}(\frac{20}{3}M_B^6 a m E_2 +
128 (1 - \frac{3 \gamma}{16}) E_4
M_B^{10}) - \tilde{\lambda}_N \tilde{\lambda}_\Sigma c_a^+ \bar{M}^2
e^{-M^2/M_B^2}
}{\tilde{\lambda_N}\tilde{\lambda_\Sigma}[2 (1-\frac{\bar{M}^2}
{M_B^2})+1]\bar{M}^2 e^{-\bar{M}^2/M_B^2}}\; ,\\
A_P &=& -\frac{\frac{A}{4^3 \pi^2 L^{4/9}}(\frac{20}{3}M_B^6 a m E_2 -
128 (1 - \frac{3 \gamma}{16} E_4
M_B^{10}) +\tilde{\lambda}_N \tilde{\lambda}_\Sigma
c_b^+ \bar{M}^2 e^{-M^2/M_B^2}
}{\tilde{\lambda_N}\tilde{\lambda_\Sigma}
\bar{M}^2 e^{-\bar{M}^2/M_B^2}} \; .
\end{eqnarray}

\begin{figure}
\begin{center}
\epsfig{file=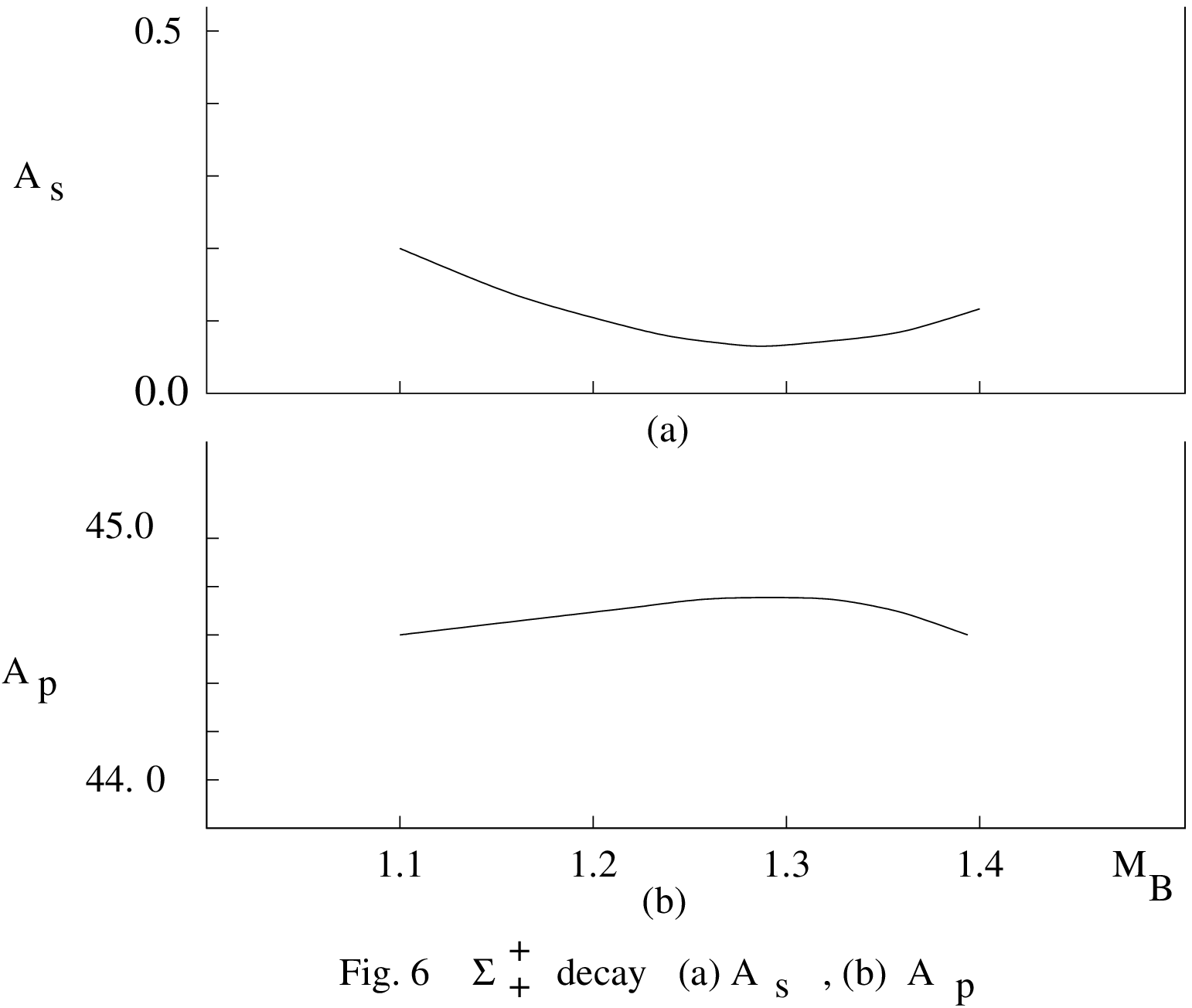,width=10cm}
%caption{}
{\label{Fig.6}}
\end{center}
\end{figure}

Experimentally, $A_S= (0.13 \pm 0.02)\times 10^{-7}, A_P =
(44.4 \pm 0.16) \times 10^{-7}, A_S/A_P \approx 0.003$. 
The single-pole parameters which give stable 
solutions for $A_s$ are $c_a^+$ = (-1.42 + 9.8 $M_B$)$\times 10^{-7}$, 
and for $A_p$ are $c_b^+$ = (-46.3 +9.5 $M_B$)$\times 10^{-7}$.The
solutions are shown in Fig.6.

\subsection {$\Sigma^-_-$ in the Pion Matrix Method}

The pion matrix method \cite {rry,SH,BK} is simpler than the external
field one 
and does not require any renormalizations. Instead of treating quarks
propagating
in an external field with a correlator defined between 
the vacuum states, the
correlator
is defined with a one-pion final state.
 This is also the starting point for light-cone
sum rules which have been used for the pion form factor\cite{braun} and
the pion wave function\cite{bj}.
One makes use of the correlator
\begin{eqnarray}
\Pi = i \int d^4x \: e^{iq\cdot x}<\pi(p=0) \mid T[ \eta_a(x) H_W
\bar{\eta}_b(0)]\mid 0> \; ,
\end{eqnarray}
rather than Eq. (9) for the external field. For the $\Sigma^-$ the 
corresponding non-vanishing diagrams are shown in Fig. 7.
Carrying out the required algebra and integrations, and doing a Borel 
transform, we obtain
%\vspace {2 in}
%Fig. 7
\begin{eqnarray}
Fig 7a&:&  \frac{4}{3} A \frac {<\bar{q} q>}{(4 \pi)^2}E_1 M_B^2
<\pi^- \mid \bar{d} i \gamma_5 u \mid 0> (\not{p} + 2m) (1 - \gamma_5) \;,
\nonumber \\
Fig. 7b&:& - \frac{4}{3} A \frac {<\bar{q} q>}{(4 \pi)^2}E_1 M_B^2
<\pi^- \mid \bar{d} i \gamma_5 u \mid 0> (\not{p} + 2m) (1 - \gamma_5) \;,\\
Fig. 7c&:& - \frac{\sqrt{2}}{3} A F_\pi \frac {1}{(4 \pi)^4}
<\pi^- \mid \bar{d} i \gamma_5 u \mid 0> (4m M_B^4 E_2 L^{-4/9} \not{p}
- 2M_B^6 E_3 L^{-4/9}) (1 + \gamma_5) \;. \nonumber
\end {eqnarray}
\begin{figure}
\begin{center}
\epsfig{file=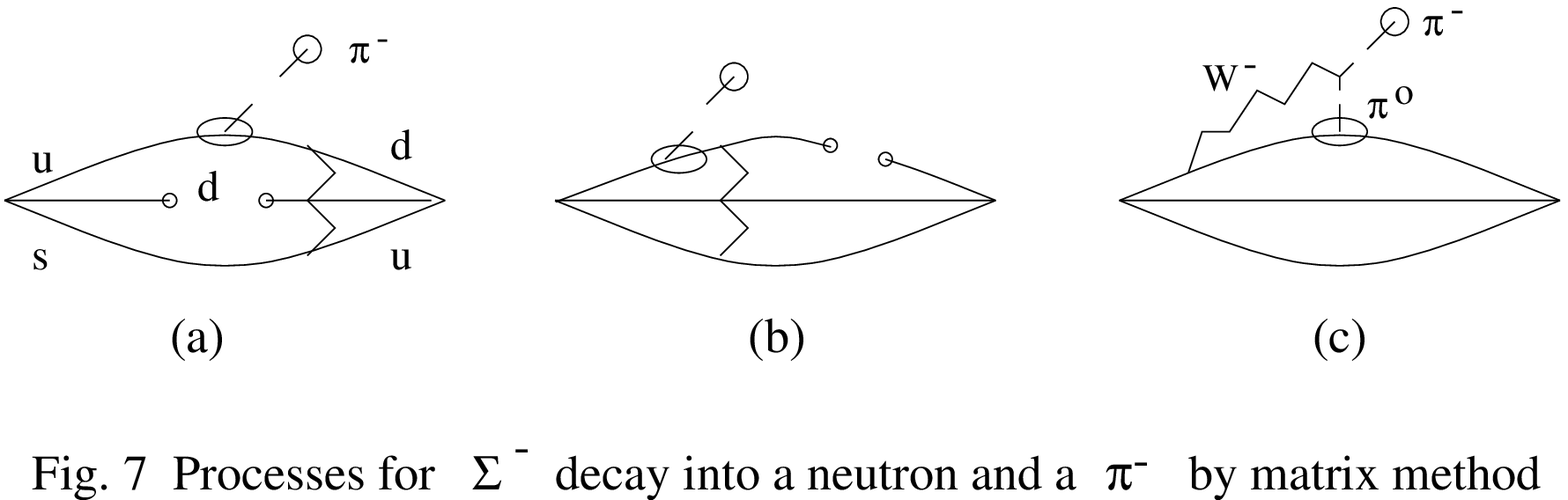,width=10cm}
%caption{}
{\label{Fig.7}}
\end{center}
\end{figure}
Note that the contributions of Figs 7a and 7b cancel exactly, so that only 
that from Fig 7c remains and it is seen that, in contrast to the
external field 
method, both S and P wave amplitudes have the same sign. The phenomenologic
side is the same as before so that we obtain
\begin{eqnarray}
\label{sum3}
A_S &=& \frac{-\frac{A}{16 \pi^2}\frac{a}{f_\pi} \frac{
M_B^8 E_3 }{L^{4/9}}+ \tilde{\lambda}_N \tilde{\lambda}_\Sigma c_a^-
 \bar{M}^2 e^{-M^2/M_B^2}}
{\tilde{\lambda}_N\tilde{\lambda}_\Sigma[2 (1-\frac{\bar{M}^2}{M_B^2})
+1]\bar{M} e^{-\bar{M}^2/M_B^2}}\; ,
\end{eqnarray}
\begin{eqnarray}
\label{sum4}
A_P &=&+ \frac{\frac{A}{16 \pi^2}\frac{a}{f_\pi} \frac{M_B^8 E_3 }{L^{4/9}}
-\tilde{\lambda}_N \tilde{\lambda}_\Sigma
c_b^-\bar{M}^2 e^{-M^2/M_B^2}}
{\tilde{\lambda}_N\tilde{\lambda}_\Sigma
  \bar{M}^2 e^{-\bar{M}^2/M_B^2}} \; ,
\end{eqnarray}
where we have put $F_\pi= 1$ and used the soft pion matrix element
\begin{eqnarray}
<0 \mid \bar{u} i \gamma_5 d \mid \pi^-> = \sqrt{2} \frac{<\bar{q} q>}{f_\pi}
= -\frac{\sqrt{2} a}{(2 \pi)^2 f_\pi},
\end{eqnarray}
with f$_\pi$ = .093 GeV in Eqs.(\ref{sum3},\ref{sum4}).

The single-pole parameters which give stable solutions
for $A_s$ are $c_a^-$ =\\
 ( -4.3 + 12.0 $M_B$) $\times 10^{-7}$, 
and for $A_p$ are $c_b^-$ = (1.13 +3.67  $M_B$)$\times 10^{-7}$.
The results are similar to those shown in Fig.4.

\subsection {$\Sigma^+_+$ Decay in the Pion Matrix Method}

%\vspace {2 in}
%Fig. 8 previous Fig. 8 dropped

\begin{figure}
\begin{center}
\epsfig{file=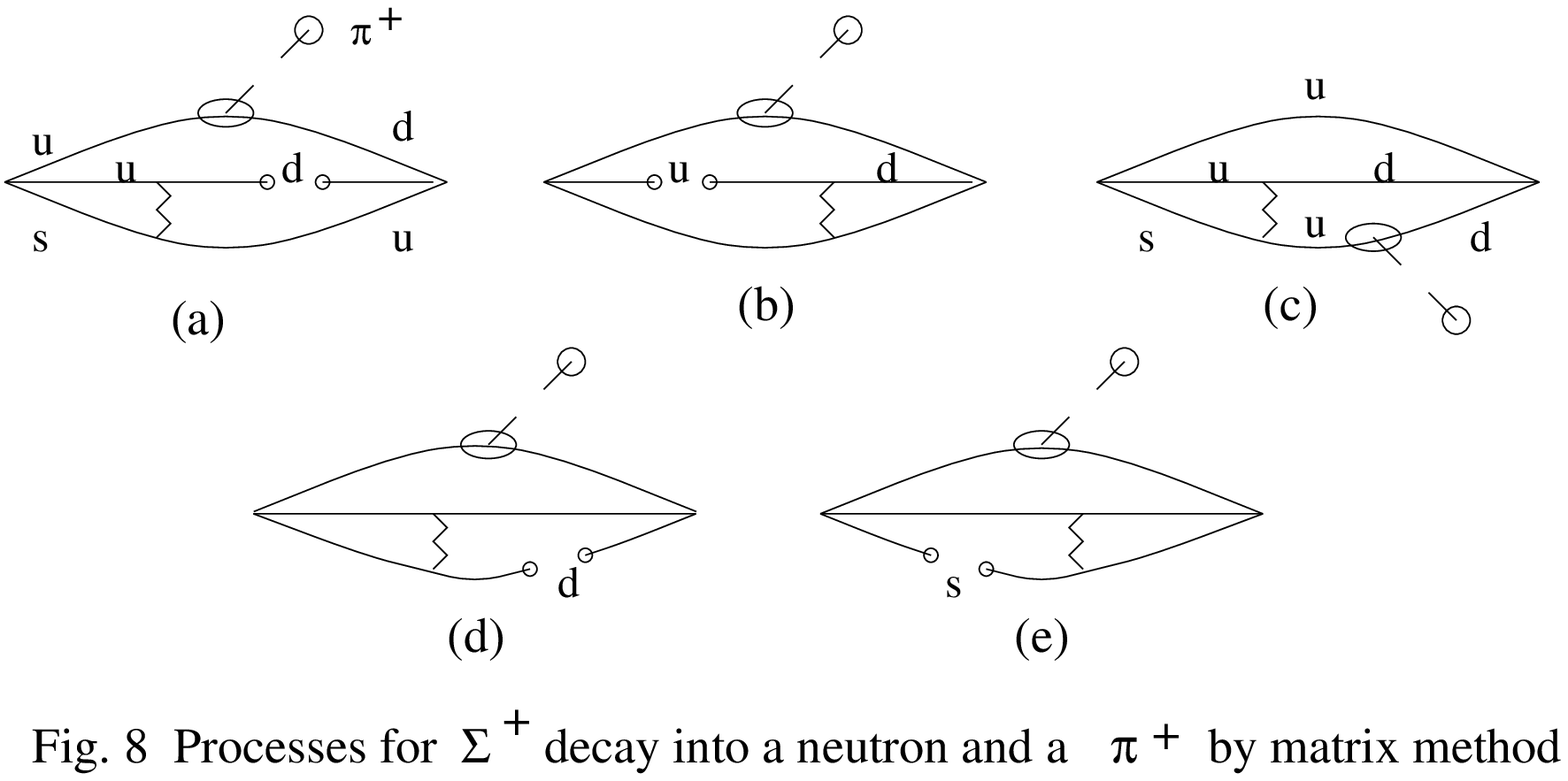,width=10cm}
%caption{}
{\label{Fig.8}}
\end{center}
\end{figure}
The contributing diagrams are shown in Fig. 8
\begin {eqnarray}
Fig 8a&:&  \frac{4}{3} A \frac {<\bar{q} q>}{(4 \pi)^2}E_1 M_B^2
<\pi^+ \mid \bar{u} i \gamma_5 d \mid 0> (\not{p} + 2m) (1 - \gamma_5) \;,
\nonumber \\
Fig.8b&:& -~ \frac{2}{3} A \frac {<\bar{q} q>}{(4 \pi)^2}E_1 M_B^2
<\pi^+ \mid \bar{u} i \gamma_5 d \mid 0> (2\not{p} + m) (1 - \gamma_5) 
\;, \nonumber\\
Fig. 8c&:& - 4 \frac{A}{(4 \pi)^4}
<\pi^+ \mid \bar{u} i \gamma_5 d \mid 0> (\frac{
4}{3} M_B^6 E_3 + m M_B^4 E_2
\not{p}\{(\frac{1}{\epsilon} - \gamma + \frac{31}{12} \nonumber \\
  & &+ [\ln{M_B^2} +(1 - \gamma)])\} (1 + \gamma_5) \;. \nonumber\\
Fig.8d&:& -~ \frac{2}{3} A \frac {<\bar{q} q>}{(4 \pi)^2}E_1 M_B^2
<\pi^+ \mid \bar{u} i \gamma_5 d \mid 0> (2\not{p} + m) (1 - \gamma_5) \;,\\
Fig.8e&:&  \frac{4}{3} A \frac {<\bar{s} s>}{(4 \pi)^2}E_1 M_B^2
<\pi^+ \mid \bar{u} i \gamma_5 d \mid 0> (\not{p} + 2m) (1 - \gamma_5) 
\;, \nonumber
\end {eqnarray}
As usual, we discard the odd sum rules; in that case no renormalization is
required. For the even sum rule we obtain
\begin {eqnarray}
\Pi =\frac{4}{3} ~\frac{A m}{(4 \pi)^2}<\pi^+ \mid \bar{u} i \gamma_5 d
\mid 0> 
M_B^2 E_1 L^{-4/9} (<\bar{q} q> 
+ 2 <\bar{s} s>) (1 - \gamma_5)\\
- \frac{16}{3} ~\frac {A}{(4 \pi)^4} <0 \mid \bar{d} i \gamma_5 u \mid \pi^+>
M_B^6 E_3 L^{-4/9}(1 + \gamma_5)
\end{eqnarray}

For $<\bar{s} s> $ we take $0.8 <\bar{q} q>$. We thus obtain for
  $A_S$ and $A_P$
\begin{eqnarray}
\label{sum5}
A_S &=&\frac{ ~\frac{\sqrt{2} A M_B^4 a}{4 \pi^2 f_\pi L^{4/9}}[
\frac{2.6}{3} m a E_1 + M_B^4 E_3 ]+ \tilde{\lambda}_N 
\tilde{\lambda}_\Sigma c_a^+ \bar{M}^2 e^{-M^2/M_B^2}}
{\tilde{\lambda_N}\tilde{\lambda_\Sigma}[2 (1-\frac{\bar{M}^2}{M_B^2})
+1]\bar{M}^2 e^{-\bar{M}^2/M_B^2}}\;
\end{eqnarray} 
\begin{eqnarray}
\label{sum6}
A_P &=&\frac{ \frac{\sqrt{2} A M_B^4 a}{4 \pi^2 f_\pi L^{4/9}}[
\frac{2.6}{3} m a E_1 - M_B^4 E_3 ]}{\tilde{\lambda_N}\tilde{\lambda_\Sigma}
\bar{M}^2 e^{-\bar{M}^2/M_B^2}} - c_b^+.
\end{eqnarray}
 
The single-pole parameters which give stable 
solutions for $A_s$ are $c_a^+$ = ( 1.0-22.3 $M_B$)$\times 10^{-7}$,
and for $A_p$ are $c_b^+$ = (-39.8 -19.7 $M_B$)$\times 10^{-7}$.
The results are similar to those shown in Fig.6.

\section{Discussion}

In a recent work by Borasoy and Holstein\cite{BH} it was found that
by including resonances in the chiral perturbation theory approach
to nonleptonic hyperon decays one can obtain fits to
data, which does not seem possible if they are not explicitly included.
In the QCD sum rule method we find it is also necessary to include 
resonances or single pole terms and have done so.

In summary, we have used QCD sum rules and two 2-point formalisms to 
examine the $\Sigma_+^+$ and $\Sigma_-^-$ nonleptonic decays.  We have assumed
that pion kinetic energy effects are small and have concentrated on the 
sum rules for which we expect these effects to be of order $\bigtriangleup
M/\bar{M} \ll 1$.We have omitted gluonic corrections, in part because we 
show that the single pole (resonance) contributions can be adjusted to fit
the data, and the changes in these parameters when gluonic corrections are 
included would not be large. Stable and consistent fits to the decay 
amplitudes are obtained with resonance contributions of similar magnitude as 
those used in chiral perturbation theory fits. 
These contributions are not known and are large, so that at the present stage
the method has no predictive power, which is also true of the chiral
perturbation theory calculations. Our results, however, parameterize quite
specific terms in the dispersion relations for the correlation functions
that we have used, and might be useful for predicting related reactions.
 
\vspace{3 mm}

\centerline{\bf Acknowledgements}
\bigskip
The work of W-Y.P.H. was supported in part by the National Science Council of
R.O.C. (NSC88-2112-M002-001Y). The work of E. M. H. was supported in part by
the U.S. Department of Energy while that of 
L.S.K. was supported in part by the National Science Foundation grant 
PHY-00070888 and in part by the U.S. D.O.E. while at the Los Alamos Natonal.
Laboratory. This work was also supported by the N.S.C. of
R.O.C. and the N.S.F. of the U.S.A. as a cooperative
research project. EMH wishes to thank the Los Alamos National Laboratory,
the Institute fuer Kernphysik at Juelich and the Theory Institute 
at Tuebingen, where some of this work was carried out and completed. In 
addition, he thanks the Humboldt Foundation for a grant. LSK wishes to thank
the TQHN group at the University of Maryland and the DOE's Institute for 
Nuclear Theory at the University of Washington for hospitality during the 
period when this work was completed.

\end{document}